\documentclass[runningheads]{llncs}
\usepackage{graphicx}
\usepackage{hyperref}

\usepackage{latexsym}
\usepackage{amssymb}
\usepackage{amsmath}
\usepackage[all]{xy}
\usepackage{lscape}
\usepackage{url}
\usepackage{adjustbox}

\usepackage{color}

\newcommand{\blue}[1]{{\color{blue} #1}}

\makeatletter
\newenvironment{prog}{\vspace{0.7ex}\par
\setlength{\parindent}{0.7cm}
\obeylines\@vobeyspaces\tt}{\vspace{0.7ex}\noindent
}
\makeatother
\newcommand{\startprog}{\begin{prog}}
\newcommand{\stopprog}{\end{prog}\noindent}

\newcommand{\alt}{\alpha} 
\newcommand{\alts}{\mathit{alts}} 
 
\newcommand{\headc}{\mathit{hd}} 
\newcommand{\inputvars}{\mathsf{var}_\mathit{in}} 
\newcommand{\outputvars}{\mathsf{out}_\mathit{vars}} 
\newcommand{\constr}{\gamma} 
\newcommand{\traces}{\mathit{Traces}} 
\newcommand{\tcases}{\mathit{TestCases}} 
\newcommand{\negcon}{\mathit{neg\_constr}} 
 
\newcommand{\rt}{\mathit{root}}

\newcommand{\id}{{\mathit{id}}} 

\newcommand{\sleq}{\leqslant}

\newcommand{\mgu}{\mathsf{mgu}}

\newcommand{\fail}{\mathsf{fail}}

\newcommand{\clauses}{\mathsf{clauses}}
\newcommand{\midd}{\!\mid\!}
\newcommand{\sep}{\mathit{\;]\![\;}}
\newcommand{\success}{\mbox{\footnotesize \textsc{success}}}
\newcommand{\failsc}{\mbox{\footnotesize \textsc{fail}}}

\def\defemb#1#2{\expandafter\def\csname #1\endcsname
                              {\relax\ifmmode #2\else\hbox{$#2$}\fi}}
\defemb{cA}{{\cal A}}
\defemb{cB}{{\cal B}}
\defemb{cC}{{\cal C}}
\defemb{cD}{{\cal D}}
\defemb{cE}{{\cal E}}
\defemb{cF}{{\cal F}}
\defemb{cG}{{\cal G}}
\defemb{cH}{{\cal H}}
\defemb{cI}{{\cal I}}
\defemb{cJ}{{\cal J}}
\defemb{cL}{{\cal L}}
\defemb{cM}{{\cal M}}
\defemb{cN}{{\cal N}}
\defemb{cO}{{\cal O}}
\defemb{cP}{{\cal P}}
\defemb{cQ}{{\cal Q}}
\defemb{cR}{{\cal R}}
\defemb{cS}{{\cal S}}
\defemb{cT}{{\cal T}}
\defemb{cU}{{\cal U}}
\defemb{cV}{{\cal V}}
\defemb{cX}{{\cal X}}
\defemb{cZ}{{\cal Z}}

\newcommand{\var}{{\cV}ar}

\renewcommand{\emptyset}{\{\}} 
\renewcommand{\phi}{\varphi}

\newcommand{\ol}[1]{\overline{#1}}  

\def \tuple#1{\langle #1 \rangle}

\newcommand{\Dom}{\mathit{Dom}}
\newcommand{\Ran}{\mathit{Ran}}

\begin{document}
\title{An SMT-Based Concolic Testing Tool\\ for Logic Programs}
\subtitle{(System Description)}
%
%

\author{Sophie Fortz \inst{1}\and Fred Mesnard \inst{2}\and  Etienne Payet \inst{2}\and Gilles Perrouin \inst{1}\and Wim Vanhoof \inst{1}\and German Vidal \inst{3}}
\authorrunning{S.~Fortz \textit{et al.}}

\institute{Universit\'e de Namur, Belgique \and %
	LIM - Universit\'e de la R\'eunion, France \and%
	MiST, DSIC, Universitat Polit\`ecnica de Val\`encia}
\maketitle              
\begin{abstract}
  Concolic testing mixes symbolic and concrete execution to generate
  test cases covering paths effectively. Its benefits have been
  demonstrated for more than 15 years to test imperative
  programs. Other programming paradigms, like logic programming, have
  received less attention. In this paper, we present a 
  concolic-based test generation method for logic programs. Our
  approach
  exploits SMT-solving for constraint resolution. We then describe the
  implementation of a concolic testing tool for Prolog and validate it on
  some selected benchmarks.

\end{abstract}

\section{Introduction}
\label{sec:intro}

Concolic testing is a well-established validation technique for
imperative and object-oriented programs 
\cite{GKS05,GZAP10,SMA05}.
However, it has been less explored in the context of
functional and logic programming languages. This is particularly
unfortunate since it is becoming increasingly popular to
\emph{compile} programs in other programming languages to either a
functional or a logic formalism. The main advantage of this approach
is that so-called declarative programs have a simpler semantics and
are thus much more appropriate for analysis, verification, validation,
etc. In particular, Constraint Horn clauses (i.e., logic programs with
constraints) are gaining popularity as an intermediate language for
verification engines (see, e.g., \cite{BMR12,GNSSS15}). In this
context, it is unfortunate that there are no powerful concolic
testing tools for logic programming languages. We thus aim at
improving this situation with the design and implementation of an
efficient SMT-based concolic testing tool for logic programs.

A notable exception in this context is the work on concolic testing
developed by Mesnard, Payet and Vidal \cite{Vid14lopstr,MPV15}, later
extended in \cite{MPV16,MPV17}. Let us illustrate this approach with a
simple example.

\begin{example}
  Consider the following logic program:
  \[
    \begin{array}{l@{~~~~}l}
      (\ell_1) & p(a).\\
      (\ell_2) & p(s(X)) \leftarrow q(X).\\
      (\ell_3)  & q(a).
    \end{array}
  \]
  where $\ell_1,\ell_2,\ell_3$ are (unique) clause labels.
  
  The notion of \emph{coverage} in \cite{MPV15} requires considering
  test cases so that the clauses defining each predicate are unified
  in all possible ways. For instance, in this example, we aim at
  producing a call to predicate $p/1$ that unifies with no clause,
  another one that unifies only with the head of  $\ell_1$, another one that
  unifies only with the head of $\ell_2$, and another one that unifies with both
  the heads of $\ell_1$ and $\ell_2$.

  For this purpose, an (iterative) process starts with an arbitrary
  test case (an atomic goal), e.g., $p(a)$. We then evaluate in
  parallel both $p(a)$ (the \emph{concrete} goal) and $p(X)$ (the
  \emph{symbolic} goal), where $X$ is a fresh variable, using a
  concolic execution extension of SLD resolution. Two key points of
  concolic execution are that not all nondeterministic executions of
  $p(X)$ are explored (but only those that corresponds to the
  executions of the concrete goal $p(a)$), and that we record the
  clauses unifying with both the concrete and the symbolic goal at
  each resolution step (in order to compute alternative test cases).

  Now, in order to look for alternative test cases, \cite{MPV15}
  introduces the notion of \emph{selective unification} problem: given
  an atom $A$, sets of clauses $H^+$ and $H^-$, and a set of variables
  $G\subseteq\var(A)$, find a substitution $\sigma$ such that i)
  $A\sigma$ unifies with the heads of the clauses in $H^+$, ii) it
  does not unify with the heads of the clauses in $H^-$, and iii)
  $G\sigma$ becomes ground.
  In this example, $p(a)$ only unifies with clause $\ell_1$, while
  $p(X)$ unifies with both $\ell_1$ and $\ell_2$. Therefore, one now
  considers the following missing cases:
  \begin{itemize}
  \item an instance $p(X)\sigma$ of $p(X)$ for some substitution
    $\sigma$ such that $p(X)\sigma$ unifies no clause, e.g., $p(b)$;
  \item another instance that unifies both $\ell_1$ and $\ell_2$
    (unfeasible if we want the argument of $p$ to be ground);
  \item and one more instance that unifies only $\ell_2$, e.g.,
    $p(s(a))$.
  \end{itemize}
  Each case is formalized as a selective unification problem and then
  solved using a specific algorithm.\footnote{More details on
    selective unification can be found in \cite{MPV16,MPV17}.}
  Now, we add the two new test cases, $p(b)$ and $p(s(a))$, and then
  repeat the process until no new test cases are added.
\end{example}
We note that, in the example above, only \emph{positive} constraints
are considered (which are denoted by means of substitutions). However,
\emph{negative} constraints (such as, e.g., $X\neq a$) cannot be
expressed in the framework of \cite{MPV15}.

Actually, the approach in \cite{MPV15} suffers from some
limitations. On the one hand, the algorithms for solving the
unification problems above are computationally very expensive, which
makes this approach impractical for large programs.  On the other
hand, the process is, in some cases, unnecessarily incomplete because
of the lack of negative information, as witnessed by the following
example:

\begin{example} \label{ex:problem}
  Consider now the following logic program:
  \[
    \begin{array}{l@{~~~~}l}
      (\ell_1) &  p(a).\\
      (\ell_2) & p(X) \leftarrow q(X).\\
      (\ell_3) & q(b).
    \end{array}
  \]
  Given the initial call $p(a)$, the first alternative test case
  computed by the approach of \cite{MPV15} is $p(b)$ which only
  unifies the second clause. Now, $p(b)$ is unfolded to $q(b)$, which
  succeeds. The next computed test case is then $p(X)\sigma$ where
  $\sigma$ binds $X$ to any term different from $b$ so that
  $q(X)\sigma$ fails (since test cases for failures are also
  required). However, it does not take into account the fact that $X$
  must be different to $a$ in $p(a)$ in order to reach $q(a)$ since
  negative constraints cannot be represented within the framework of
  \cite{MPV15}.  Hence, one could generate $p(a)$ so that concolic
  testing stops because $p(a)$ was already considered. The generated
  test cases are then $p(a)$ and $p(b)$. With the intended coverage
  definition, though, one would also expect a test case like $p(c)$
  which is first unfolded using the second clause and then
  fails. Therefore, the concolic testing framework of \cite{MPV15} is
  unnecessarily incomplete here.
\end{example}
In this paper, we design an improved concolic testing scheme that is
based on \cite{MPV15} but adds support for negative constraints (so
that the source of incompleteness shown in the example above is
removed) and defines selective unification problems as constraints on
Herbrand terms, so that an efficient SMT solver can be used (in
contrast to \cite{MPV16,MPV17}). We have implemented an SMT-based
concolic testing tool for Prolog based on this design, where the SMT
solver Z3 \cite{Z3DeMoura2008} is used to solve selective unification
problems. A preliminary experimental evaluation has been conducted,
which shows encouraging results in terms of execution time and scalability.

\section{A Deterministic Operational Semantics}
\label{sec:background}

In this section, we recall a \emph{deterministic} operational
semantics for \emph{definite} logic programs (i.e., logic programs
without negation \cite{Llo87}). In particular, we consider the
semantics in \cite{MPV15} which, in turn, follows the \emph{local}
operational semantics of \cite{SESGF11}, where backtracking is dealt
with explicitly.
Moreover, the semantics only considers the computation of the first
answer for the initial goal. This is a design decision motivated by
the fact that Prolog programs are often used in this way, so that one
can measure the achieved coverage in a realistic way.

We refer the reader to \cite{Apt97} for the standard definitions and
notations for logic programs. The semantics is defined by means of a
transition system on \emph{states} of the form
$\tuple{\cB^1_{\delta_1}\midd\ldots\midd\cB^n_{\delta_n}}$, where
$\cB^1_{\delta_1}\midd\ldots\midd\cB^n_{\delta_n}$ is a sequence of
goals labeled with substitutions (the answer computed so far, when
restricted to the variables of the initial goal).  We denote sequences
with $S, S',\ldots$, where $\epsilon$ denotes the empty sequence.  In
some cases, we label a goal $\cB$ both with a substitution and a
program clause, e.g., $\cB_\delta^{H\leftarrow\cB}$, which is used to
determine the next clause to be used for an SLD resolution step (see
rules \textsf{choice} and \textsf{unfold} in
Figure~\ref{fig:concrete2}).  Note that the clauses of the program are
not included in the state but considered as global parameters since
they are static.
In the following, given an atom $A$
and a logic program $P$, $\clauses(A,P)$
returns the sequence of renamed apart program clauses $c_1,\ldots,c_n$ 
from $P$ whose head unifies with $A$.
A syntactic object $s_1$ is \emph{more general} than a syntactic object
$s_2$, denoted $s_1 \sleq s_2$, if there exists a substitution
$\theta$ such that $s_1\theta = s_2$.
$\var(o)$ denotes the set of variables of the syntactic object $o$. 
For a substitution $\theta$, $\var(\theta)$ is defined as
$\Dom(\theta)\cup\Ran(\theta)$, where $\Dom$ and $\Ran$ return the
variables in the domain and range of a given substitution, respectively.

\begin{figure}[t]
  \rule{\linewidth}{1pt}
  \[
  \hspace{-2ex}\begin{array}{r@{~}l}
    \mathsf{(success)} & {\displaystyle 
      \frac{~} 
        {\tuple{\mathsf{true}_\delta\midd S} \to \tuple{\success_\delta}}
        } 
        \\[2ex]

     \mathsf{(failure)} & {\displaystyle 
      \frac{~} 
        {\tuple{(\fail,\cB)_\delta} \to \tuple{\failsc_\delta}}
        }
        \hspace{24ex}
    \mathsf{(backtrack)} ~ {\displaystyle 
      \frac{S\neq\epsilon} 
        {\tuple{(\fail,\cB)_\delta\midd S} \to \tuple{S}}
        }\\[4ex]

     \textsf{(choice)} &  {\displaystyle 
      \frac{
        \clauses(A,P) = (c_1,\ldots,c_n)\wedge n>0 
      } 
        {\tuple{(A,\cB)_\delta\midd S} \to
          \tuple{(A,\cB)_\delta^{c_1} \midd \ldots\midd
          (A,\cB)_\delta^{c_n}\midd  S}}
        } 
        \hfill\hspace{1ex}
    \mathsf{(choice\_fail)} ~ {\displaystyle 
      \frac{
        \clauses(A,P)=\emptyset} 
        {\tuple{(A,\cB)_\delta\midd S} \to \tuple{(\fail,\cB)_\delta\midd S}}
        }\\[3ex]

    \mathsf{(unfold)} & {\displaystyle 
      \frac{\mgu(A,H_1)=\sigma} 
        {\tuple{(A,\cB)_\delta^{H_1\leftarrow\cB_1}\midd S}
          \to \tuple{(\cB_1\sigma,\cB\sigma)_{\delta\sigma}\midd S}}
        } \\ 

    \end{array}
    \]
  \rule{\linewidth}{1pt}
  \caption{Concrete semantics} 
  \label{fig:concrete2}
\end{figure}

For simplicity, w.l.o.g., we only consider \emph{atomic} initial
goals. Therefore, given an atom $A$, an initial state has the form
$\tuple{A_\id}$, where $\id$ denotes the identity substitution.
The transition rules, shown in Figure~\ref{fig:concrete2}, proceed as
follows:
\begin{itemize}
\item In rules \textsf{success} and \textsf{failure}, we use constant
  $\success_\delta$ to denote that a successful derivation ended with
  computed answer substitution $\delta$, while $\failsc_\delta$
  denotes a finitely failing derivation; recording $\delta$ for
  failing computations might be useful for debugging purposes.

\item Rule \textsf{backtrack} applies when the first goal in the
  sequence finitely fails, but there is at least one alternative
  choice. 

\item Rule \textsf{choice} represents the first stage of an SLD
  resolution step. If there is at least one clause whose head unifies
  with the leftmost atom, this rule introduces as many copies of a
  goal as clauses returned by function $\clauses$. If there is at
  least one matching clause, unfolding is then performed by rule
  \textsf{unfold}. Otherwise, if there is no matching clause, rule
  \textsf{choice\_fail} returns \textsf{fail} so that either rule
  \textsf{failure} or \textsf{backtrack} applies next.

\end{itemize}

\begin{example} \label{ex1}
  Consider the following logic program:
  \[
  \begin{array}{l@{~~~~~~}l@{~~~~~~}l}
    p(a). &
    q(a).   &
    r(a).  \\
  
    p(s(X)) \leftarrow q(X). & q(b). & r(c). \\
    
    p(f(X)) \leftarrow r(X). & \\
    \end{array}
    \]
    Given the initial goal $p(s(X))$, we have the following successful
    computation (for clarity, we label each step with the applied
    rule):\footnote{Note that a fact like ``$q(a).$'' is equivalent to
      a rule ``$q(a)\leftarrow \mathsf{true}.$'' and, thus, the
      unfolding of $q(X)$ returns $\mathsf{true}$ in the considered
      derivation.}
    \[
    \begin{array}{llllll}
      \tuple{p(s(X))_\id} & \to^\mathsf{choice} &
      \tuple{p(s(X))_\id^{p(s(X)) \leftarrow q(X)}} & \to^\mathsf{unfold}
      & \tuple{q(X)_\id} \\
      & \to^\mathsf{choice} & \tuple{q(X)_\id^{q(a)}\midd q(X)_\id^{q(b)}}
      & \to^\mathsf{unfold}
      & \tuple{\mathsf{true}_{\{X/a\}}\midd q(X)_\id^{q(b)}} \\
      & \to^\mathsf{success} & \tuple{\success_{\{X/a\}}} 
    \end{array}
    \]
    Therefore, we have a successful computation for $p(s(X))$ with
    computed answer $\{X/a\}$. Observe that only the first answer is
    considered.
\end{example}

\section{An SMT-Based Concolic Testing Procedure}
\label{sec:approach}

In this section, we present our scheme to concolic testing of logic
programs.
%
Our concolic testing semantics performs both concolic execution and
test case generation, which contrasts to \cite{MPV15} where they are
kept as two independent stages. Moreover, it collects both positive
and negative constraints on the input variables, so that it overcomes
an important limitation of previous approaches (as explained in
Section~\ref{sec:intro}) and opens the door to making the overall
process much more efficient by using a state-of-the-art SMT solver.

\subsection{Auxiliary Functions and Notations}

Let us first introduce some auxiliary definitions and notations
which are required in the remainder of this section.

In the following, we let $\ol{o_n}$ denote the sequence of syntactic
objects $o_1,\ldots,o_n$; we also write $\ol{o}$ when the number of
elements is not relevant. Given an atom $A$, we let $\rt(A)= p/n$ if
$A= p(\ol{t_n})$. We also assume that every clause $c$ has a
corresponding unique label, which we denote by $\ell(c)$. By abuse of
notation, we also denote by $\ell(\ol{c_n})$ the set of labels
$\{\ell(c_1),\ldots,\ell(c_n)\}$. Moreover, we let
$\headc(H\leftarrow \cB) =
H$ 
and $\headc(C) = \{\headc(c)\mid c\in C\}$, where $H\leftarrow \cB$ is
a clause and $C$ is a set of clauses.

We now introduce the auxiliary functions $\negcon$ and $\alts$. First,
function $\negcon$ is used to compute some negative constraints which
will become useful to avoid the problem shown in
Example~\ref{ex:problem}. 

\begin{definition}[$\negcon$]
  Let $A$ be an atom, $G\subseteq\var(A)$ a set of variables, and
  $\{\ol{H_n}\}$, $n>0$, a set of atoms with
  $\var(\{\ol{H_n}\})\cap\var(A) =\emptyset$. Then,
  \[
    \negcon(A, \{\ol{H_n}\},G) = 
    \forall \ol{X_{k_1}} A\neq H_1\wedge\ldots\wedge \forall
    \ol{X_{k_n}} A\neq H_n
  \]
  where 
  $\ol{X_{k_i}} = (\var(A)\setminus G) \cup\var(H_i)$, $i=1,\ldots,n$.
\end{definition}
For example, we have
\[
  \negcon(p(X,Y),\{p(a,W)\},\{X\}) = \forall Y,\! W\: p(X,Y)\neq p(a,W)
\]
Function $\alts$ is used to encode a selective unification problem
using both positive and negative constraints:
Intuitively, given a call of the form
$\alts(A_0,\constr,A',\cB,\cB',G)$, we are interested in new test
cases that unify with each atom in each set of
$\cP(\cB')$\footnote{Here, we denote by $\cP(C)$ the powerset of a set
  $C$.} except for the set $\cB$ which is already \emph{covered} by
the current concrete goal (i.e., we do not want to produce an
alternative test case that matches exactly the same clauses as the
current test case that we are executing).  Then, for each set
$H^+\in \cP(\cB')$ (the \emph{positive} clauses) with
$H^- = \cB'\setminus H^+$ (the \emph{negative} clauses), we look for a
substitution $\sigma$ such that $A'\sigma$ unifies with the atoms in
$H^+$ but it does not unify with the atoms in $H^-$, while still
grounding the variables in $G$. For each such substitution, we produce
a new test case $A_0\sigma$. Formally,

\begin{definition}[$\alts$]
  Let $A_0,A'$ be atoms, $\cB,\cB'$ sets of atoms with
  $\var(\cB,\cB')\cap\var(A')=\emptyset$, $\constr$ a (negative)
  constraint, and $G\subseteq\var(A_0)$ a set of variables. Then, we
  have
  \[\def\arraystretch{1.5}
    \alts(A_0,\constr,A',\cB,\cB',G) = \left\{ A_0\sigma
      \;\middle|\;
  \begin{array}{l}
    H^+\in \cP(\cB'),\ H^+\neq \cB, \\
    H^- = \cB' \setminus H^+,\\
    \alt(A',\constr,H^+,H^-,G) = \sigma
  \end{array}\right\}\]
  where
  function $\alt(A',\constr,H^+,H^-,G)$ returns a solution to the
  following constraint (represented as a
  substitution):\footnote{Function $\alt$ returns an arbitrary
    solution when the considered constraint is satisfiable and fails
    otherwise.} 
    \[\def\arraystretch{1.5}
  \left(
  \begin{array}{llll}
    \constr & \wedge & \exists \ol{X_{n_1}} (A' = H_1) \wedge
    \ldots \wedge \exists \ol{X_{n_j}} (A'=H_j) \\
    & \wedge & \forall \ol{Y_{n_1}} (A' \neq H'_1) \wedge \ldots
    \wedge \forall \ol{Y_{n_k}} (A' \neq H'_k)
  \end{array}\right)\]
  with $H^+=\{\ol{H_j}\}$,
  $H^-=\{\ol{H'_k}\}$,
  $\ol{X_{n_i}} = (\var(A')\setminus G)\cup \var(H_i)$,
  $i=1,\ldots,j$, and
  $\ol{Y_{n_i}} = (\var(A')\setminus G)\cup \var(H'_i)$,
  $i=1,\ldots,k$. Here, we look for a solution for the free variables
  of the above constraint: $\var(A') \cap G$.
\end{definition}
For example, we have
\[
  \alts(p(X),p(X)\neq p(a),q(X),\emptyset,\{q(b)\},\{X\})
  = \{ p(c) \}
\]
since
\[
  \alt(q(X),p(X)\neq p(a), \emptyset,\{q(b)\},\{X\}) =
  \exists X(p(X)\neq p(a)\wedge q(X)\neq q(b))
\]
and the selected solution is $X=c$ represented as a substitution
$\{X/c\}$, where $c$ is an arbitrary constant which is different from
the previous ones ($a$ and $b$).

\subsection{Concolic Testing Semantics}
\label{subsection:ConcolicTestingSemantics}

Let us now consider our concolic testing semantics. 
In this work, we consider that the initial goal is terminating for a
given \emph{mode}, which is a reasonable assumption since
non-terminating test cases are not very useful in practice.
Essentially, a mode is a function that labels as ``input'' or
``output'' the arguments of a given predicate, so that input arguments
are assumed to be ground at call time, while output arguments are
usually unbound (see, e.g., \cite{Apt97}).  Here, we assume a fixed
\emph{mode} for the predicate in the initial (atomic) goal. Different
modes can also be considered by performing concolic testing once for
each mode of interest.
Therefore, our test
cases should make the input arguments ground (to ensure terminating
concrete executions). For clarity, we assume in the remainder of this
paper that all input arguments (if any) are in the first consecutive
positions of an atom. Moreover, given a predicate $p/n$ with input
arguments $\{1,\ldots,m\}$, $m\leq n$, we let
$\inputvars(p(\ol{t_n}))=\var(t_1)\cup\ldots\cup\var(t_m)$, i.e., the
variables in the input positions, and
$\outputvars(p(\ol{t_n})) = p(\ol{t_{m}},X_{m+1},\ldots,X_n)$, i.e.,
the atom that results from replacing the output arguments by distinct
fresh variables.

In the concolic testing semantics, \emph{concolic states} have now the
form $\tuple{S\sep S'}$, where $S$ and $S'$ are sequences of (possibly
labeled) concrete and symbolic goals, respectively. In the context of
logic programming, the notion of \emph{symbolic} execution is very
natural: the structure of both $S$ and $S'$ is the same, and the only
difference (besides the labeling) is that some atoms might be less
instantiated in $S'$ than in $S$.
To be precise, symbolic goals in a concolic state are now labeled as
follows: $\cB_{\sigma,\pi,A_0,\gamma,G}$, where
\begin{itemize}
\item $\sigma$ is the substitution computed so far;
\item $\pi$ is the current trace\footnote{Traces are sequences of
    clause labels, with $\epsilon$ the empty trace.} (which is required
  to avoid considering the same computations once and again);
\item $A_0$ is the initial atomic goal (which is required to compute
  alternative test cases in function $\alts$);
\item $\gamma$ is a (negative) constraint that allows us to avoid
  unification with the heads of some clauses (to avoid the problem
  shown in Example~\ref{ex:problem});
\item $G$ is the set of variables that must be ground in test cases
  (i.e., the variables in the \emph{input} arguments of the initial
  atomic goal).
\end{itemize} 
%
%
Our concolic testing semantics considers two \emph{global parameters}
that are left implicit in the transition rules: $\traces$ and
$\tcases$. The first one, $\traces$, is used to store the already
explored execution traces, so that we avoid computing the same test
cases once and again. Here, a \emph{trace} represents a particular
execution by the sequence of clause labels used in the unfolding steps
of this execution.

The second one, $\tcases$, stores the computed
test cases, where the already processed test cases are distinguished
by underlining them (i.e., $\underline{A}$ means that the test case
$A$ has been already processed by the concolic testing semantics).

Concolic testing consists of an iterative process where $\tcases$ is
initialized with some arbitrary (atomic) goal, e.g.,
$\tcases = \{ A \}$, $\traces$ is initialized to the empty set, and it
proceeds as follows:
\[
  \begin{array}{l}
    \mathbf{repeat}\\
    \hspace{3ex} \mathbf{let}~ p(\ol{t_n},\ol{X_m}) \in\tcases\\
    \hspace{3ex} \tcases \leftarrow (\tcases - \{p(\ol{t_n},\ol{X_m})\}) \cup
    \{\underline{p(\ol{t_n},\ol{X_m})}\}\\
    \hspace{3ex} \mathbf{execute}~ \tuple{p(\ol{t_n},\ol{X_m})_\id
    \sep p(\ol{Y_{n+m}})_{\id,\epsilon,p(\ol{Y_{n+m}}),\mathit{true},\{\ol{Y_n}\}}} \\
    \mathbf{until}~\mbox{all atoms in $\tcases$ are underlined} 
  \end{array}
\]
where we assume that the first $n$ arguments of $p$ are its input
arguments, and concolic states are executed using the semantics in
Figure~\ref{fig:concolic} (see below).

In general, though, this iterative process might run forever, even if
all considered concolic executions are terminating.\footnote{In
  principle, we assume that concrete goals are terminating and, thus,
  concolic executions are terminating too.} Essentially, our algorithm
aims at full \emph{path coverage}, so the required number of test
cases is typically infinite.  Therefore, in practice, one usually sets
a bound in the number of iterations, a time limit, or a maximum term
depth so that the domains of terms and atoms become finite. In the
implemented tool, we consider the last approach (see
Section~\ref{sec:implem}).

\begin{figure}[t]
  \rule{\linewidth}{1pt}\\[-1ex]
  \[
  \hspace{-4ex}\begin{array}{r@{~}l}
    \mathsf{(success)} & {\displaystyle 
      \frac{~~} 
        {\tuple{\mathsf{true}_{\delta}\midd S
            \sep\mathsf{true}_{\theta,\pi,A_0,\constr,G}\midd S'} 
          \leadsto
          \tuple{\success_{\delta} \sep \success_{\theta}}}
        } \\[4ex]

    \mathsf{(failure)} & {\displaystyle 
      \frac{~} 
        {\tuple{(\mathsf{fail},\cB)_{\delta}
          \sep (\mathsf{fail},\cB')_{\theta,\pi,A_0,\constr,G}} \leadsto 
        \tuple{\failsc_{\delta} \sep \failsc_{\theta}}}
        }\\[4ex]

    \mathsf{(backtrack)} & {\displaystyle 
      \frac{S\neq\epsilon} 
        {\tuple{(\mathsf{fail},\cB)_{\delta}\midd S \sep
          (\mathsf{fail},\cB')_{\theta,\pi,A_0,\constr,G}\midd S'} \leadsto \tuple{S\sep S'}}
        }\\[4ex]

     \mathsf{(choice)} &  {\displaystyle 
                         \frac{\begin{array}{l}
                                 \clauses(A,P) = \ol{c_n}\wedge n>0
                                 \wedge \clauses(A',P)=\ol{d_k}\\
                                 \hspace{12ex}\wedge\; \constr' =
                                 \negcon(A',\headc(\{\ol{d_k}\})\setminus
                                 \headc(\{\ol{c_n}\}),G)
                                 \end{array}
                                 } 
        {\begin{array}{l@{~}l}
            \tuple{(A,\cB)_{\delta}\midd S \sep (A',\cB')_{\theta,\pi,A_0,\constr,G}\midd S'} \\
            \hspace{6ex}\leadsto 
          \tuple{(A,\cB)_{\delta}^{c_1} \midd \ldots\midd
           (A,\cB)_{\delta}^{c_n}\midd  S \\
           \hspace{9ex} \sep (A',\cB')_{\theta, \pi.\ell(c_1) ,A_0,\constr\wedge\constr',G}^{c_1} \midd \ldots\midd
          (A',\cB')_{\theta, \pi.\ell(c_n) ,A_0,\constr\wedge\constr',G}^{c_n}\midd  S'
        }\end{array}}
                 }\\[8ex]
                       & \blue{\mbox{where}}\\
                 &\blue{
                         \begin{array}{lll}
				~~~\tcases \leftarrow\tcases\cup\outputvars(\alts(A_0,\constr,A',\headc(\{\ol{c_n}\}),\headc(\{\ol{d_k}\}),G)) \\
				~~~\traces \leftarrow \traces \cup\{\pi\} \\
                         \end{array} 
                 }\\
                 &\blue{\mbox{if}~\pi\not\in\traces}\\[2ex]

    \mathsf{(choice\_fail)} & {\displaystyle 
      \frac{\clauses(A,P)=\emptyset
              \wedge \clauses(A',P)=\ol{d_k}\wedge \constr'= \negcon(A',\headc(\{\ol{d_k}\}),G)} 
        {\tuple{(A,\cB)_{\delta}\midd S \sep (A',\cB')_{\theta,\pi,A_0,\constr,G}\midd S'} 
          \leadsto
          \tuple{(\fail,\cB)_{\delta}\midd S\sep (\fail,\cB')_{\theta,\pi,A_0,\constr\wedge\constr',G}\midd S'}}
              }\\[2ex]
                       & \blue{\mbox{where}}\\
                 &\blue{
                         \begin{array}{lll}
				~~~\tcases \leftarrow\tcases\cup\outputvars(\alts(A_0,\constr,A',\emptyset,\headc(\{\ol{d_k}\}),G)) \\
				~~~\traces \leftarrow \traces \cup\{\pi\} \\
                         \end{array} 
                 }\\
                 &\blue{\mbox{if}~\pi\not\in\traces}\\[2ex]

    \mathsf{(unfold)} & {\displaystyle 
      \frac{\mgu(A,H_1)=\sigma\wedge\mgu(A',H_1)=\rho} 
                        {\begin{array}{l}
                           \tuple{(A,\cB)_{\delta}^{H_1\leftarrow\cB_1}\midd S\sep
          (A',\cB')_{\theta,\pi,A_0,\constr,G}^{H_1\leftarrow\cB_1}\midd S'}\\
          \leadsto \tuple{(\cB_1\sigma,\cB\sigma)_{\delta\sigma}\midd S
          \sep
                           (\cB_1\rho,\cB'\rho)_{\theta\rho,\pi,A_0\rho,\constr\rho,
                           \var(G\rho)}\midd S'}
                           \end{array}}
        } \\ 

    \end{array}
    \]
  \rule{\linewidth}{1pt}
  \caption{Concolic testing semantics} \label{fig:concolic}
\end{figure}

Finally, let us briefly describe the rules of the concolic testing
semantics shown in Figure~\ref{fig:concolic}:

Rules $\mathsf{success}$, $\mathsf{failure}$, $\mathsf{backtrack}$,
and $\mathsf{unfold}$ are straightforward extensions of the same rules
in the concrete operational semantics of Figure~\ref{fig:concrete2}.

As for rules $\mathsf{choice}$ and $\mathsf{choice\_fail}$, if we only
look at the first component of concolic states, they are identical to
their counterpart in Figure~\ref{fig:concrete2}; indeed, the concolic
testing semantics is a conservative extension of the standard
operational semantics.
Regarding the symbolic components, there are several notable
differences:
\begin{itemize}
\item First, although the symbolic goal is only unfolded using the
  clauses matching with the concrete goal, we also determine the set
  of clauses matching the symbolic goal, $\ol{d_k}$. This information
  will be useful in order to compute alternative test goals (using the
  auxiliary function $\alts$).

\item Moreover, we update the current trace (from $\pi$ to
  $\pi.\ell(c_i)$ in rule $\mathsf{choice}$) and the negative
  constraint from $\gamma$ to $\gamma\wedge\gamma'$, where $\gamma'$
  is used to ensure that the symbolic goal, $A'$, only matches the
  same clauses as the concrete goal $A$.

\item Finally, observe how the global parameters $\tcases$ and
  $\traces$ are updated in these rules when $\pi\not\in\traces$, i.e.,
  when the considered execution path is considered for the first time
  (otherwise, we just continue the concolic execution without
  modifying $\tcases$ nor $\traces$). Note that we use $\outputvars$
  here to ensure that the output arguments are unbound.
\end{itemize}

\subsection{Concolic Testing in Practice}

Consider again the logic program from Example~\ref{ex:problem}:
\[
  \begin{array}{l@{~~~~}l}
    (\ell_1) &  p(a).\\
    (\ell_2) & p(X) \leftarrow q(X).\\
    (\ell_3) & q(b).
  \end{array}
\]
and the initial call $p(a)$, so that we initialize $\tcases$ to
$\{p(a)\}$ and, thus, start concolic testing with
$ \tuple{p(a)_\id \sep p(Y)_{\id,\epsilon,p(Y),\mathit{true},\{Y\}}}
$, where we assume that the argument of $p/1$ is an input argument and
it is expected to be ground. Concolic testing then computes the
following derivation:\footnote{In this example, we often use clause
  labels instead of the actual clauses for clarity.}
\[
  \begin{array}{lll}
    \tuple{p(a)_\id
    \sep p(Y)_{\id,\epsilon,p(Y),\mathit{true},\{Y\}}} \\
    \leadsto^\mathsf{choice}
    \tuple{p(a)_\id^{\ell_1} \midd p(a)_\id^{\ell_2}
    \sep p(Y)_{\id,\ell_1,p(Y),\mathit{true},\{Y\}}^{\ell_1}
    \midd p(Y)_{\id,\ell_2,p(Y),\mathit{true},\{Y\}}^{\ell_2}} \\
    \leadsto^\mathsf{unfold}
    \tuple{\mathsf{true}_\id \midd p(a)_\id^{\ell_2}
    \sep \mathsf{true}_{\id,\ell_1,p(Y),\mathit{true},\{Y\}}
    \midd p(Y)_{\id,\ell_2,p(Y),\mathit{true},\{Y\}}^{\ell_2}} \\
    \leadsto^\mathsf{success}
    \tuple{\success_\id \sep \success_\id}
  \end{array}
\]
Moreover, in the first step, we add $\epsilon$ to $\traces$ and update
$\tcases$ as follows:
\[
  \begin{array}{lll}      
    \tcases & \leftarrow & \tcases \\
            & \cup &
                     \outputvars(\alts(p(Y),\mathit{true},p(Y),\{p(a),p(X)\},\{p(a),p(X)\},\{Y\}))
  \end{array}
\]
In this case,
$\cP(\{p(a),p(X)\}) = \{\emptyset,\{p(a)\},\{p(X)\},\{p(a),p(X)\}\}$,
and we exclude the last element since this case is already
considered. Therefore, following the definition of function $\alts$,
we consider the following three candidates for $H^+$ in order to
compute alternative test cases:
  \begin{itemize}
  \item $H^+ = \emptyset$: here, we have $H^- = \{p(a),p(X)\}$
    and we should find a solution to 
    $\alt(p(Y),\mathit{true},\emptyset,\{p(a),p(X)\},\{Y\})$,
    i.e., find a ground instance of $p(Y)$ that unifies with no
    clause. In particular, one must solve the following constraint:
    \[
      \exists Y (\mathit{true}\wedge p(Y)\neq p(a) \wedge \forall X
      \:p(Y)\neq p(X))
    \]
    which is trivially unfeasible.

  \item $H^+=\{p(a)\}$: here, we have $H^- = \{p(X)\}$ and we should
    find a solution to
    $\alt(p(Y),\mathit{true},\{p(a)\},\{p(X)\},\{Y\})$, i.e.,
    find a ground instance of $p(Y)$ that only unifies with the
    first clause. In this case, one must solve the following
    constraint:
    \[
      \exists Y (\mathit{true} \wedge p(Y)= p(a) \wedge \forall X
      \:p(Y)\neq p(X))
    \]
    which is also unfeasible.

  \item $H^+ = \{p(X)\}$: here, we have $H^- = \{p(a)\}$ and we should
    find a solution to
    $\alt(p(Y),\mathit{true},\{p(X)\},\{p(a)\},\{Y\})$, i.e.,
    find a ground instance of $p(Y)$ that only unifies with the
    second clause. In this case, one must solve the following
    constraint:
    \[
      \exists Y (\mathit{true}\wedge \exists X\:p(Y)= p(X) \wedge 
      p(Y)\neq p(a))
    \]
    which is satisfiable with model, e.g., $X=b,Y=b$ (represented by
    the substitution $\{X/b,Y/b\}$). Therefore, we add a new test case
    $p(b)$ to $\tcases$.
  \end{itemize}
  In the second iteration of the concolic testing algorithm, we
  consider the initial goal $p(b)$ and thus, we start concolic testing
  with
  $ \tuple{p(b)_\id \sep p(Y)_{\id,\epsilon,p(Y),\mathit{true},\{Y\}}}
  $. Concolic testing then computes then the following
  derivation:
  \[
    \begin{array}{lll}
      \tuple{p(b)_\id
      \sep p(Y)_{\id,\epsilon,p(Y),\mathit{true},\{Y\}}} \\
                                                             \hspace{20ex}\leadsto^\mathsf{choice}
       \tuple{p(b)_\id^{\ell_2} \sep
      p(Y)_{\id,\ell_2,p(Y),p(Y)\neq p(a),\{Y\}}^{\ell_2}} \\
                                                             \hspace{20ex}\leadsto^\mathsf{unfold}
       \tuple{q(b)_\id 
      \sep q(X)_{\{Y/X\},\ell_2,p(X),p(X)\neq p(a),\{X\}}} \\
                                                                   \hspace{20ex}\leadsto^\mathsf{choice}
       \tuple{q(b)_\id^{\ell_3} \sep
      q(X)_{\{Y/X\},\ell_2.\ell_3,p(X),p(X)\neq p(a),\{X\}}^{\ell_3}} \\
                                                             \hspace{20ex}\leadsto^\mathsf{unfold}
       \tuple{\mathsf{true}_\id 
      \sep \mathsf{true}_{\{Y/b\},\ell_2.\ell_3,p(b),p(b)\neq p(a),\emptyset}} \\
      \hspace{20ex}\leadsto^\mathsf{success}
      \tuple{\success_\id \sep \success_{\{Y/b\}}}
    \end{array}
  \]
  Moreover, in the second choice step, we add $\ell_2$ to $\traces$
  and update $\tcases$ as follows:
  \[
    \begin{array}{lll}
      \tcases & \leftarrow & \tcases \\
              & \cup &
                       \outputvars(\alts(p(X),p(X)\neq
                       p(a),q(X),\{q(b)\},\{q(b)\},\{X\}))
  \end{array}
  \]
  Now, we have $\cP(\{q(b)\}) = \{\emptyset,\{q(b)\}\}$, and we
  exclude the last element since this case is already
  considered. Therefore, there is only one candidate for computing
  alternative test cases (i.e., $H^+=\emptyset$) and we should find a
  solution to
  \[
    \alt(q(X),p(X)\neq p(a),\emptyset,\{q(b)\},\{X\})
  \]
  i.e., finding a ground instance of $q(X)$ that does not unify with
  $q(b)$ and, moreover, $p(X)\neq p(a)$ holds. In particular, one must
  solve the following constraint:
  \[
    \exists X (p(X)\neq p(a) \wedge q(X)\neq q(b))
  \]
  which is trivially feasible.\footnote{Here, we assume that the
    considered domain includes at least one more constant, e.g., $c$.}
  Note that the solution $X=a$ is now ruled out thanks to the negative
  constraint $p(X)\neq p(a)$, thus avoiding the problem shown in
  Example~\ref{ex:problem}.  A possible solution is, e.g., $X=c$, so
  that the next test case that we add to $\tcases$ is $p(c)$.

  In the third (and last) iteration, no more alternatives are obtained
  (we omit the concolic testing derivation for brevity), so our
  algorithm produces three test cases for this example: $p(a)$, $p(b)$
  and $p(c)$, achieving a full coverage.


\section{Implementation and Experimental Evaluation}
\label{sec:implem}

In this section, we describe the implementation of an SMT-based
concolic testing tool for Prolog programs that follows the ideas
presented so far.

\begin{figure}[t]
  \includegraphics[width=\linewidth]{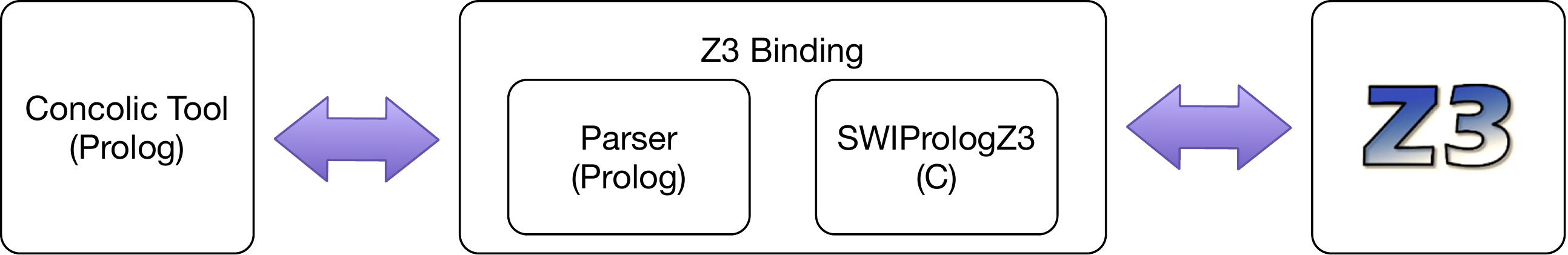}
  \caption{Implementation Workflow}
  \label{fig:implem}
\end{figure}

We have implemented our tool in SWI-Prolog \cite{WSTL12} and have used
the C interface of SWI-Prolog in order to call the functions of the Z3
solver \cite{Z3DeMoura2008}. The scheme of the workflow is shown in
Figure~\ref{fig:implem}.
Here, ``Concolic Tool'' is the main module (written in Prolog) which
performs concolic execution. Once a constraint is built, it is
transformed into an SMT well-formed string and sent to the module
``SWIPrologZ3'', a module written in C that uses Z3's library to
interact with the Z3 solver and solve this constraint. The
results are sent back to the main module ``Concolic Tool''.

In contrast to the concolic testing semantics shown in
Figure~\ref{fig:concolic}, we have implemented a
\emph{nondeterministic} version of concolic execution which is based
on Prolog's backtracking mechanism. Here, the information that must
survive a backtracking step is inserted to the internal database using
dynamic predicates and asserted for consistency.

Regarding the termination of concolic testing, we impose a maximum
term depth for the generated test cases. Since the domain is finite
and we do not generate duplicated test cases, termination is trivially
ensured. Consider, for instance, the usual specification of natural
numbers built from $0$ and $s(\_)$:
\[
  \begin{array}{r@{~}l}
    & (\ell_1) ~nat(0).\\ 
    & (\ell_2)~nat(s(X)) \leftarrow nat(X).
  \end{array}
\]
where we assume that the argument of $nat/1$ is an \emph{input}
argument (and must be ground). Given an initial goal like $nat(0)$, in
the first iteration of the algorithm we add, e.g., the following new
test cases: $nat(1)$ and $nat(s(0))$, where the constant $1$ is used
to avoid matching any clause. When considering the second test case,
we will generate the alternative test cases $nat(s(1))$ and
$nat(s(s(0))$. And the process goes on forever.

In this context, by setting a maximum term depth, e.g., $2$, one can
limit the generated test cases to only
\[
  \{nat(0),nat(1),nat(s(0)),nat(s(1)),nat(s(s(0))),nat(s(s(1)))\}
\]
The maximum term depth is an input parameter of our concolic testing
tool since it depends on the particular program and the desired
code coverage.

Finally, we show some selected results from a preliminary experimental
evaluation of our concolic testing tool. We aimed at addressing
the following questions:
\begin{enumerate}
\item {\sf Q1: What is the performance of our technique on typical
    benchmarks?} Here, the goal was to assess the viability of the
  proposed method by measuring its execution time on some selected
  benchmarks.
  
\item {\sf Q2: How does it compare to existing tools for concolic
    testing?} In particular, we wanted to consider the tool
  \textsf{contest} \cite{MPV15}, which is publicly available through a
  web interface.\footnote{Moreover, a copy of the Prolog sources of
    \textsf{contest} were provided by its authors.}
\end{enumerate}

\paragraph{Benchmarks.} We selected six subject programs from
previous benchmarks \cite{MPV15} and from
GitHub.\footnote{\url{https://github.com/Anniepoo/prolog-examples}}
We ran concolic testing between 3 and 100 executions on a MacBook Pro
hexacore 2,6 Ghz with 16 GB RAM in order to get reliable results.
Reported times, in seconds, are the average of these executions. Our results are
reported in Table~\ref{table:results}. Here, \textsf{concolic} refers
to the tool presented in this paper, while \textsf{contest} refers to
the tool introduced in \cite{MPV15}. The \textsf{size} of a subject program is the number of its source lines of code. The
column \textsf{Ground Args} displays the number of arguments
of the initial symbolic goal 
to ground, starting at the first position.
\textsf{\#TCs}
refers to the number of generated test cases. 
A timeout for \textsf{contest} is set to 1000 seconds (the
\textsf{crash} is an overflow).

\begin{table}[t]
\caption{Summary of experimental results} \label{table:results}
\hspace{-2ex}
  \begin{tabular}{|l|l|l|l|l|l|l|l|l|} \hline
    \textsf{Subject} & & & \textsf{Ground} & \textsf{Max} & \textsf{time} & \textsf{time} &
                                                              \textsf{\#TCs}
    & \textsf{\#TCs}\\
    \textsf{program} & \textsf{size} & \textsf{Initial goal} &
       \textsf{Args} & \textsf{Depth} & \textsf{concolic} &
\textsf{contest} & \textsf{concolic} & \textsf{contest} \\\hline\hline
Nat                                                       & 2    & nat(0)                                                                                       & 1                                                              & 1     & 0.050                                                           & 0.0273                                                         & 3                                                             & 4                                                            \\ \hline
Nat                                                     & 2    & nat(0)                                                                                       & 1                                                              & 5     & 0.0897                                                          & 0.1554                                                         & 7                                                             & 12                                                           \\ \hline
Nat                                                   & 2    & nat(0)                                                                                       & 1                                                              & 50    & 1.6752                                                          & 19.5678                                                        & 52                                                            & 102                                                          \\ \hline
Generator                                                           & 7    & generate(empty,\_A,\_B)                                                                      & 1                                                              & 1     & 1.4517                                                          & 0.7096                                                         & 9                                                             & 9                                                            \\ \hline
Generator                                                              & 7    & generate(empty,T,\_B)                                                                        & 2                                                              & 1     & 1.3255                                                          & 4.4820                                                         & 9                                                             & 9                                                            \\ \hline
Generator                                                               & 7    & generate(empty,T,H)                                                                          & 3                                                              & 1     & 1.3211                                                          & \textsf{crash}                                              & 9                                                             & N/A                                                          \\ \hline
Activities                                                              & 38   & \begin{tabular}[c]{@{}l@{}}what\_to\_do\_today(sunday,\\ sunny,wash\_your\_car)\end{tabular} & 3                                                              & 2     & 6.3257                                                          & \textsf{timeout}                               & 122                                                           & N/A                                                          \\ \hline
Cannibals  & 78   & start(config(3,3,0,0))                                                                       & 1                                                              & 2     & 0.0535                                                          & \textsf{timeout}                                 & 2                                                             & N/A                                                          \\ \hline
Family                                                           & 48   & parent(dicky,X)                                                                            & 1                                                              & 1     & 20.0305                                                         & 64.1838                                                        & 9                                                             & 19                                                           \\ \hline
Monsters &&&&&&&&\\
and mazes & 113  & base\_score(will,grace)
                         & 2
                                           & 2     & 0.2001
                                                             & 0.4701
                                                                             &
                                                                               6                                                             & 7     \\\hline
\end{tabular}
\end{table}

\subsubsection{\sf Q1: Performance.}

The three first lines of Table \ref{table:results} clearly show the
influence of the maximum term depth in a typically recursive
program. Our procedure handles recursion better than \textsf{contest}
because we end up generating complex constraints that are more
efficiently solved using an SMT solver than using the specific
algorithms in \cite{MPV15}, as expected. 
For simpler cases, though, interacting with the solver is likely
more expensive than performing the computations in \cite{MPV15}.
However, as the complexity increases, 
our SMT-based technique is faster and scales better than
\textsf{contest}.
These preliminary experiments support our choice of using a powerful
SMT solver for test case generation.

\subsubsection{\sf Q2: Comparison to other concolic testing tools.}

Besides scalability, which is already considered above, we noticed
that our tool typically produces less test cases that
\textsf{contest}. In principle, this can be explained by the fact that the
algorithm in \cite{MPV15} allows one to also bind the \emph{output}
arguments of the initial goal. Consider, e.g., the following simple
program:
\[
  \begin{array}{l}
    p(a,b).\\
    p(X,c).\\
  \end{array}
\]
where the first argument is an input argument and the second one is an
output argument.  Here, \textsf{contest} might return four test cases:
\begin{itemize}
\item $p(a,a)$, which matches no clause;
\item $p(a,b)$, which only matches the first clause;
\item $p(a,c)$, which only matches the second clause;
\item $p(a,Y)$, which matches both clauses.
\end{itemize}
In contrast, our tool would only return the test cases $p(a,Y)$, which
matches both clauses, and $p(b,Y)$, which only matches the second
clause, since the second argument is an output argument and so, it cannot be bound.


This is essentially a design decision, but we think it is more
sensible to keep output arguments unbound in test cases.

On the other hand, we also noticed that, in some cases, our tool
produced some test cases that were not generated by
\textsf{contest}. This is explained by the problem illustrated in
Example~\ref{ex:problem}. 



\paragraph{Threats to Validity.}

These experiments are preliminary and are therefore subject to
validity threats.  We mitigated internal validity by repeating our
experiments several times and ensuring the validity of the produced
test cases manually.  We alleviated external validity by selecting
programs of varying size publicly available on GitHub, though we
cannot guarantee they are representative.

\section{Conclusion}
\label{sec:conclusion}

In this paper, we have designed an improved concolic testing scheme
that is based on \cite{MPV15} but adds support for negative
constraints and defines \emph{selective unification problems} as
constraints on Herbrand terms. Our approach overcomes some
of the problems in previous approaches \cite{MPV16,MPV17}, mainly
regarding the scalability of the technique as well as a potential
source of incompleteness due to the lack of negative information.
We have implemented an SMT-based concolic testing tool in SWI-Prolog
that uses the Z3 library for C through the foreign language interface
of SWI-Prolog. Our preliminary experimental evaluation has shown
promising results regarding the scalability of the method in
comparison to previous approaches.

As for future work, we plan to extend and improve our
concolic testing tool. In particular, we will consider its extension to CLP programs \cite{JM94}. Given the way selective unification problems are represented in this paper (as
constraint satisfiability problems), dealing with
constraints over domains other than Herbrand terms seems very natural.
Finally, we will formally prove the correctness and
completeness of the improved algorithms.

%

\bibliographystyle{splncs04}
\bibliography{biblio}

\end{document}